\documentclass[12pt,letterpaper,hyper]{JHEP3}
\usepackage{amsmath,epsfig}
\usepackage{amssymb,amsfonts}
\usepackage{graphicx}
\usepackage{subfigure}

\newcommand{\bi}{\begin{itemize}}
\newcommand{\ei}{\end{itemize}}
\newcommand{\non}{\nonumber}
\def\p{\partial}

\def\b{\beta}
\def\d{\delta}
\def\g{\gamma}
\def\l{\lambda}

\def\k{\kappa}
\def\s{\sigma}

\def\O{\mathcal{O}}

\def\G{\Gamma}
\def\D{\Delta}

\def\O{\mathcal{O}}

\def\om{\omega}
\def\r{\rightarrow}

\def\half{{\frac12}}

\newcommand{\bea}{\begin{eqnarray}}
\newcommand{\eea}{\end{eqnarray}}
\newcommand{\be}{\begin{equation}}
\newcommand{\ee}{\end{equation}}

\title{Exploring the BTZ bulk with boundary conformal blocks}

\author{Bruno Carneiro da Cunha$^\dag$ and Monica Guica$^{\flat,\sharp}$ \\

\hspace{-4.5mm}{\small $ {}^ \dag$ Departamento de F\'isica, Universidade Federal de Pernambuco,
\\ \hspace{-0.35 cm}  50670-901, Recife, Pernambuco, Brazil}

\vskip2mm

\hspace{-4.5mm}{\small $ {}^\flat$   Nordita, Stockholm University and KTH Royal Institute of Technology,\\
\hspace{-0.35 cm}
Roslagstullsbacken 23, SE-106 91 Stockholm, Sweden
}
\vskip2mm

\hspace{-4.6mm}{\small $ {}^\sharp$   Department of Physics and Astronomy, Uppsala University,\\  \hspace{-0.35 cm}  SE-751 08 Uppsala, Sweden\\
}}
\abstract{ 

\vskip5mm

We  point out a simple relation between the bulk field at an arbitrary radial position and the boundary OPE, by placing some old work by Ferrara, Gatto, Grillo and Parisi in the AdS/CFT context. This gives us, in principle, a prescription for extracting the classical bulk field from the  boundary conformal block, and also clarifies why the latter is computed by a geodesic Witten diagram. We apply this prescription to the BTZ black hole - viewed as a pure state created by the insertion of a heavy operator in the boundary CFT$_2$ - and use it to relate a classical  field in the bulk to a heavy-light Virasoro conformal block in the boundary. In particular, we obtain a relation between the radial bulk position and the conformal ratios in the boundary CFT. We use this to show that   the singular points of the radial bulk  equation occur when the dual boundary operators approach each other and that the associated bulk monodromies map to monodromies of the  (appropriately transformed) conformal block, thus  providing a CFT interpretation of the radial monodromy. 

}

\begin{document}

\section{Introduction}

One of the most basic tools for exploring the internal structure of black holes is via scattering of light fields.  For example, scattering in a black hole background readily reveals holographic features, such as the well-known connection between  greybody factors and thermal CFT Green's functions \cite{Maldacena:1996ix,Maldacena:1997ih}.  Interesting applications include tests of linear stability of solutions in pure general relativity (see e.g. \cite{Berti:2009kk} and references therein) and relaxation times in AdS/CFT \cite{Horowitz:1999jd}. More recent results connect black hole scattering to chaos \cite{Maldacena:2015waa,Shenker:2013pqa}  and to braiding relations in the dual CFT \cite{Jackson:2014nla}.
  
One particular, powerful method to study black hole scattering is the monodromy method \cite{Motl:2003cd,Neitzke:2003mz,Castro:2013lba}, which uses the analytic properties of the wave equation in the complexified radial plane to determine the scattering data. The method can generically be  applied to a wide class of black hole solutions; in particular, scalar perturbations of four dimensional Kerr \cite{daCunha:2015ana} and Kerr-de Sitter \cite{daCunha:2015fna} black holes have their scattering coefficients given implicitly in terms of Painlev\'e transcendents\footnote{ These  transcendents functions have been related to conformal blocks in CFT$_2$  in \cite{Gamayun:2012ma}.} .

 Quite universally, the radial wave equation has singular points at the black hole horizons and infinity, with associated monodromies that encode the thermodynamic properties of the black hole and suggest  a dual CFT$_2$-like interpretation of generic black holes \cite{Castro:2013kea}. 
%
However, the precise significance of the singular points and their associated monodromies is not understood from the dual boundary perspective even in the simple case of the BTZ black hole, where the 
holographic dual is exactly a CFT$_2$.

In order to understand the CFT significance of the singular points, one needs an expression for the bulk field at some arbitrary radial position $r$ in terms of purely boundary quantities. The best-known  CFT representation of the bulk field is given by the HKLL prescription \cite{Banks:1998dd,Hamilton:2005ju,Hamilton:2006az,Hamilton:2006fh,Heemskerk:2012mn}, which uses the bulk-to-boundary Green's function in the background of interest to express an operator in the interior of AdS in terms of its boundary values. This construction is clearly background - and thus state - dependent (there is further state dependence if one passes behind the black hole's  horizon \cite{Papadodimas:2012aq,Papadodimas:2013jku}); moreover, since it takes the bulk solution as an input, this prescription cannot be used to extract the meaning of the radial direction from the CFT perspective.  

 The main purpose of this paper is to propose an interpretation of the bulk field in BTZ
that mostly uses CFT input, leading to an identification between the radial plane in the bulk and the plane of the boundary CFT, as well as  a relation between the classical bulk field and the boundary conformal block. We then  use this  to map the  singular points of the bulk wave equation to situations where the boundary operators approach each other, and to relate the associated bulk and boundary  monodromies. 

 Our starting point is a relation between the bulk field and the boundary OPE that first appeared in the works of Ferrara, Gatto, Grillo and Parisi 
 \cite{Ferrara:1971vh,fgg} (without the AdS interpretation) and was  indirectly used in the recent proposal \cite{Hijano:2015zsa} that conformal blocks are computed by geodesic  Witten diagrams. Basically, \cite{Ferrara:1971vh}  found it most convenient  to express the contribution of the conformal family of a primary operator $\O$ of dimension $\D$ to the OPE of two scalar operators $A,B$   as 
\bea
A(x) B(0) &\sim & \frac{\mathcal{B}^{-1}_{AB}}{|x|^{\D_A +\D_B} } \int_0^1 \frac{du}{u(1-u)} \left(\frac{u}{1-u}\right)^{\frac{\D_{AB}}{2}} \G(\nu+1) 2^\nu \, \times \non \\
&& \times \int \frac{d^d p}{(2\pi)^d}\, \frac{ e^{i u p\cdot  x}}{(-p^2)^\frac{\nu}{2}} \,  \left(u(1-u) x^2\right)^{\frac{d}{4}} \, J_\nu (\sqrt{- u(1-u) x^2 p^2})\,  \O(p)\label{fggope}
\eea
where $ \nu = \Delta - \frac{d}{2}$, $\D_{AB} = \D_A -\D_B$, $\mathcal{B}_{AB}= \mathcal{B}\left(\frac{\D + \D_{AB} }{2}, \frac{\D - \D_{AB}}{2}\right)$ is the Euler beta function of the respective arguments.  We have set the fusion coefficient $C_{AB}^\O =1$ and we have slightly generalised the formula of \cite{Ferrara:1971vh} to arbitrary CFT dimension, $d$.
For $x^2$ spacelike and $p^2$ timelike, one immediately recognises the expression under the integral to be the free bulk field $\Phi^{(0)}$ corresponding to  the boundary operator $\O$,  at position 

\be
z(u) = \sqrt{u(1-u) x^2} \;, \;\;\;\;\;\; x^\mu(u) = u \, x^\mu  \label{adsgeo}
\ee
in Poincar\'e AdS$_{d+1}$. This path in AdS  traces the bulk geodesic $\g_{AB}$ that joins the insertions of  the operators $A$ and $B$ in the boundary theory.  Introducing  a new parameter $\l$ such that $u = (1+e^{2\l})^{-1}$, the above relation can be written more suggestively as 

\be
A(x) B(0) \sim \frac{2\mathcal{B}^{-1}_{AB}}{|x|^{\D_A +\D_B}} \int_{-\infty}^{\infty} d\l \, e^{-\l \D_{AB}}\, \Phi^{(0)} (y(\l)) \label{fggrels}
\ee
where  $y(\l) = (z(\l),x(\l))$ is the bulk position evaluted along the geodesic $\g_{AB}$. It should then not be surprising that the four-point global conformal block  simply corresponds to the bulk-to-bulk propagator $\langle \Phi^{(0)} \Phi^{(0)}  \rangle$ with endpoints  integrated over the geodesics pairwise uniting the positions  of the external operators,  yielding the geodesic Witten diagram prescription of \cite{Hijano:2015zsa}. 

According to the above relation, the bulk field is simply a convenient way to encode the conformal family of the primary operator $\O$ that appears in the OPE of two boundary operators\footnote{This is particularly clear in CFT$_1$, where \eqref{fggrels} is replaced by the even simpler relation \eqref{2dbfope}.}.  The conformal Casimir equation satisfied by $\O$ and its descendants is translated into  the wave equation satisfied by the bulk field.  Note that  \eqref{fggrels}  is simply a kinematical statement, which follows from from the conformal symmetry of the CFT vacuum, and does not use any dynamical CFT information. That is, in any CFT we can write the conformal family of an operator appearing in the OPE of two other operators in terms of a free bulk field in an auxiliary AdS spacetime; however, this bulk field will in general not be causal. Only when the CFT has a large $N$ expansion will the causal bulk field  be well approximated by a free one, as the multitrace contributions that are needed to restore causality \cite{Kabat:2011rz} are suppressed at leading order in $1/N$.

Given \eqref{fggrels}, one may try to ``undo'' the OPE  of the two boundary operators  to obtain the bulk field. More precisely, note that \eqref{fggrels}, upon the substitution $t = e^{-\l}$, represents the OPE of the boundary operators $A,B$ as the Mellin transform  of the free bulk field with respect to $\D_{AB}$. Thus we can define, at least formally, the bulk field  evaluated along a boundary-anchored geodesic as the inverse Mellin transform of the OPE with respect to $\D_{AB}$. Note that it does not matter which external operators $A,B$ we use to define the bulk field, as long as the fusion coefficient $C_{AB}^\O \neq 0$; the dependence on $\D_A+\D_B$ is only in the overall prefactor, whereas the dependence on $\D_{AB}$ is done away via the inverse Mellin transform.  
This inverse Mellin transform operation may be best justified when it acts inside correlation functions,  which can be viewed as  functions of the external operator dimensions.  It is also interesting to note that  while the Mellin transform of the CFT correlators with respect to the  spatial coordinates effectively lives in the space of conformal dimensions\footnote{Remember the constraint on the Mellin parameters is $\sum_j \delta_{ij} = \D_i$ \cite{Fitzpatrick:2011ia,Paulos:2011ie}.}, here the inverse Mellin transform replaces a relative conformal dimension  by an additional spatial coordinate, which can be taken to be the AdS geodesic parameter, $\l$. This parameter allows us to reach into the bulk,  with the maximum bulk penetration (obtained for $\l=0$) fixed by the separation of the operators $A,B$ in the CFT. Thus, we obtain a relation between the radial depth in the bulk and the spatial separation in the CFT plane.

The advantage of relating the bulk field to the boundary OPE is that this relation will hold inside arbitrary correlation functions. For example, the correlator $\langle \Phi^{(0)} AB \rangle$ will be related to the conformal block $\langle A B \, \mathbb{P}_\O  C D \rangle$ via a Mellin transform with respect to $\D_{CD}$. This expectation value represents  a classical bulk field  that is sourced on the geodesic uniting the boundary insertions of $C,D$ \cite{Hijano:2015zsa}. As argued in \cite{Hijano:2015zsa},  performing a geodesic rather than a bulk integral  effectively projects the correlator onto the conformal family of a specific operator.

What we have  discussed so far is an alternate way to understand the bulk field in \emph{vacuum} AdS, which gives the same result as  the HKLL prescription. However, what we are ultimately interested in is to understand the bulk field in non-trivial backgrounds; we will concentrate on the BTZ black hole \cite{Banados:1992wn} for the purposes of this paper.  Following e.g. \cite{Jackson:2014nla,Verlinde:2015qfa}, our model for the BTZ black hole will be that of a pure state created by acting with a heavy CFT operator (i.e., whose (anti)holomorphic dimensions scale as $h_H, \bar h_H \sim c$) on the CFT vacuum.  As argued in a  nice series of papers \cite{Fitzpatrick:2014vua,Fitzpatrick:2015zha}, correlation functions of light operators in the state created by the heavy one behave as thermal correlators, characterised by the left/right  temperatures\footnote{The actual temperature is given by $T_H = 2 T_L T_R /(T_L + T_R)$.} 
\be
T_{L} = \frac{1}{2 \pi} \sqrt{\frac{24 h_H}{c} - 1} \;, \;\;\;\;\;\;\;\;T_{R} = \frac{1}{2 \pi} \sqrt{\frac{24 \bar h_H}{ c} - 1} \label{temp}
\ee
 The gravitational dual of this statement is that the heavy operators backreact on the geometry, creating a BTZ black hole, which is then probed by the light operators. Given that the BTZ black hole is locally equivalent to pure AdS$_3$ \cite{Banados:1992gq}, the free bulk field in this background is simply given by a coordinate transformation of the bulk field sourced on a geodesic in vacuum AdS$_3$, and can be interpreted as the bulk field sourced on a ``heavy'', backreacted geodesic \cite{Hijano:2015qja}. The same authors show that geodesic Witten diagrams where one of the two geodesics has been backreacted  compute heavy-light Virasoro conformal blocks in the dual CFT$_2$.

 
All this suggests that we can obtain the free bulk field in BTZ by undoing the
integral along the ``light'' geodesic in the geodesic Witten diagram, which should correspond to undoing the  OPE of the light operators inside the heavy-light Virasoro conformal block. This procedure appears to be   state-independent (the inverse Mellin transform acting on the light operators is the same, no matter which heavy operators we choose to create the black hole state), but it produces a bulk field that is explicitly state-dependent, as it encodes all the couplings of the light operators and boundary gravitons to the heavy operators. 
 
It may not be immediately clear why the bulk field obtained via this procedure satisfies the wave equation in BTZ. After all, the Virasoro block, when rewritten in terms of global conformal blocks, consists of an infinite sum over multitrace operators constructed from the exchanged operator $\O$ and various powers of the stress tensor. If the wave equation in the bulk is to correspond to the Casimir equation on the boundary, we note that each of the exchanged global blocks satisfies a Casimir equation  with a different eigenvalue.  However, as  shown in \cite{ Fitzpatrick:2015zha}, the heavy-light Virasoro block reduces to the global conformal block for $\O$ in a conformally transformed background, $w$. Thus, in the new coordinates, the entire Virasoro block satisfies a  Casimir equation with invariant $\D(\D-2)$, which is nothing but the bulk  wave equation. 
A similar identification between the boundary Casimir and the bulk wave equation has  previously appeared in \cite{Verlinde:2015qfa}, but in the context of a somewhat different construction. 

The relation between the bulk field and the heavy-light conformal block yields a new identification between the bulk radial coordinate and the dual CFT conformal cross-ratios. This allows us to find a CFT interpretation of the singular points of the wave equation - which now occur when operators in the block approach each other - and of the associated monodromies. 

This paper is organised as follows. In section \ref{bfcb} we review the results of \cite{fgg} and discuss the relation between the bulk field, the boundary OPE and geodesic Witten diagrams in \emph{vacuum} AdS. In section \ref{bfbtz} we discuss bulk fields in BTZ and their relation to heavy-light Virasoro conformal blocks, in light of the work of \cite{ Fitzpatrick:2015zha}. In section \ref{cfti}, we map the singular points of the wave equation in BTZ to special values of the dual CFT conformal ratios and relate the monodromies of the bulk field with those of the boundary conformal block. We conclude with some discussion. In appendix \ref{cbdet}, we reproduce some details of computation of conformal blocks first presented in \cite{fgg}.

As this article was nearing completion, \cite{Czech:2016xec} appeared, which has some overlap with  section \ref{bfcb}.

\section{Bulk fields and conformal blocks \label{bfcb}}

In this section, we briefly review the work of \cite{Ferrara:1971vh,fgg} and  relate it to the recent work \cite{Hijano:2015zsa}, outlining how the bulk field operator and the geodesic Witten diagram appear in their expressions for the boundary OPE and  the four-point conformal block, respectively.

\subsection{Bulk fields from the boundary OPE \label{bfbope}}

The best known way to construct the (normalizable) bulk field operator in AdS in terms of operators in the boundary CFT is the so-called HKLL prescription\footnote{The HKLL bulk field also receives multitrace contributions that are subleading in $1/N$ \cite{Kabat:2011rz}. The superscript $^{(0)}$ on $\Phi^{(0)}$ indicates that we are only considering the leading contribution, i.e. the exactly free bulk field. }

\be
\Phi^{(0)} (z,x) = \int d^d x' K(z,x| x') \O(x') \label{kabat}
\ee
where $K(z,x| x')$ satisfies the free wave equation in the bulk ($z,x^\mu$ are the usual Poincar\'e coordinates) and its normalization is fixed such that

\be
\lim_{z \r 0} z^{-\Delta} \Phi^{(0)}(z,x) = \O(x)
\ee
This expression \eqref{kabat} is best understood by Fourier transforming with respect to the boundary coordinates, case in which \cite{Hamilton:2006az}
\be
\Phi^{(0)}(z,x) = 2^\nu \Gamma(\nu+1) \int \frac{d^d p}{(2\pi)^d} \frac{e^{i p \cdot x}}{(-p^2)^{\frac{\nu}{2}}} z^{\frac{d}{2}} J_\nu (z \sqrt{-p^2} ) \O(p)
\ee
In the introduction we remarked the striking similarity between this expression for the bulk field and the integrand  appearing in the formula \eqref{fggrels} for the OPE of two scalar operators. The latter  was
 obtained in \cite{Ferrara:1971vh} via the usual procedure of fixing the descendant terms in the OPE expansion such that the end result matches with the three-point function, whose form is entirely fixed by conformal symmetry. \cite{Ferrara:1971vh} found it useful to express the three-point function in terms of a Schwinger parameter, $u$, which then also appears in the OPE, and has the interpretation of geodesic parameter in AdS.

The relation \eqref{fggrels} leads us to interpret  the bulk field as the conformal family of $\O$ that appears in the OPE of two other primary operators. Note that this interpretation only holds  when the operators are \emph{spacelike separated}. When the boundary points are not spacelike separated, one cannot extract the bulk field from the OPE, as the corresponding field in AdS would have spacelike momentum, and would thus be unphysical.
In the case of CFT$_1$, where the boundary operators can only be timelike separated, \eqref{fggrels} is replaced by an even simpler relation between the bulk field and the boundary OPE, which we will now derive. Incidentally, this CFT$_1$ relation also makes the clearest the conformal family interpretation of the bulk field.

Consider the boundary representation of a bulk scalar in AdS$_2$. We want the bulk field to be normalizable at infinity, so we are working in \emph{Lorentzian} signature. According to \cite{Hamilton:2005ju}, we have
\be
\Phi^{(0)} (z, t) = \frac{\G (\D + \half)}{\sqrt{\pi} \G(\D)}\int_{-\infty}^\infty dt' \left( \frac{z^2 - (t-t')^2}{z} \right)^{\Delta-1} \Theta(z-|t-t'|)\, \O(t') \label{kabads2}
\ee 
The $\Theta$ function is  important in making the integral converge. Setting $t =0$ for convenience and letting $t' = \eta\, z$ with $\eta \in (-1,1)$, the integral becomes

\bea
\Phi^{(0)}  (z, 0) & = & z^\Delta  \,\frac{\G (\D + \half)}{\sqrt{\pi} \G(\D)} \int_{-1}^1 d\eta \, (1-\eta^2)^{\D-1} \sum_n \frac{z^n \eta^n}{n!} \p^n \O(0) \non \\
&=&   z^\Delta \,\frac{\G (\D + \half)}{\sqrt{\pi} \G(\D)} \sum_n \frac{z^n}{n!} \frac{(1+(-1)^n)\G(\D) \G(\frac{1+n}{2})}{2 \G(\frac{n+1}{2} +\D)} \, \p^n \O(0)\non \\
&=&  z^\Delta \,\frac{\G (\D + \half)}{\sqrt{\pi} } \sum_k \frac{z^{2k}}{(2k)!} \frac{ \G(k+\half)}{ \G(k +\D + \half)} \, \p^{2k} \O(0) \label{expphi2}
\eea
It is interesting to note that this is related to the descendants of $\O$ that appear in the OPE of two scalars $A,B$ of \emph{equal} dimension $\D_A = \D_B$

\be
A(t_1) B(t_2) \sim \frac{1}{(2\D t)^{2\D_A-\D}} \sum_n c_n (\D t)^n \p^n \O(\bar t) \;, \;\;\;\;\;\;\;\; \D t = \frac{t_2-t_1}{2} \;, \;\;\;\;\bar t = \frac{t_1+t_2}{2}
\ee
where the $c_n$ are fixed by matching with the three-point function (we set $c_0 =1$) and read\footnote{We  used the identity $\G(z)\G(z+\half) = 2^{1-2z} \sqrt{\pi} \, \G(2z)$.}
%

\be
c_{2k} = \frac{1}{k!} \frac{\G(\D+k) \G(2\D)}{\G(\D)\Gamma(2\D + 2 k)} = \frac{1}{ (2k)! \sqrt{\pi}} \frac{\G(\D + \half) \G(k+\half)}{\G(\D +k+\half)}\;, \;\;\;\;\;\;\; c_{2k+1} =0
\ee
These are identical to the coefficients in \eqref{expphi2}, so we can write
\be
A(t_1) B(t_2) \sim \frac{2^\D}{(2\D t)^{2\D_A}} \, \Phi^{(0)} (|\Delta t|,\bar t) \label{2dbfope}
\ee
Thus, in one dimension, the conformal descendants that appear in the OPE of two equal-dimension scalars are exactly encoded in an AdS$_2$ bulk field at radial position $z = |\D t|$. This implies e.g. that the four-point conformal block with pairwise equal external operator dimensions is equal to the bulk two-point function, a relation that can easily be checked.

%
%
%

\subsection{Correlators and geodesic Witten diagrams \label{crrgwd}}

The relation \eqref{fggrels} between the bulk field and the boundary OPE implies that we can use geodesic Witten diagrams to represent certain bulk or boundary correlators. We will be particularly interested in the three-point correlator
\be
\Phi_{\g_{\!{}_{AB}}} (y)  \equiv \langle \Phi^{(0)} (y) A (x_1) B (x_2) \rangle
\ee
where $y$ denotes a bulk point and $x^i$ denote boundary points. 
Using \eqref{fggrels}, this can be reduced to 

\be
\Phi_{\g_{\!{}_{AB}}} (y) = \frac{2\,\mathcal{B}_{AB}^{-1}}{|x_{12}|^{\D_A +\D_B}} \int_{\g_{\!{}_{AB}}} d\l \, e^{-\l \D_{AB}} \langle  \Phi^{(0)} (y) \Phi^{(0)} (y'(\l))\rangle
\ee
which tells us that $\Phi_{\g_{\!{}_{AB}}} (y)$ is a solution to the free bulk wave equation, with a $\d$-function source on the geodesic $\g_{\!{}_{AB}}$.  Given that $\langle \Phi^{(0)} \Phi^{(0)} \rangle $ equals the bulk-to-bulk propagator $G_{bb}$,  the above correlator is computed by the following geodesic Witten diagram

\bigskip

\medskip

\begin{figure}[h]
  \begin{minipage}{0.45\textwidth}
  \begin{flushright}
    \includegraphics[height=2.8cm]{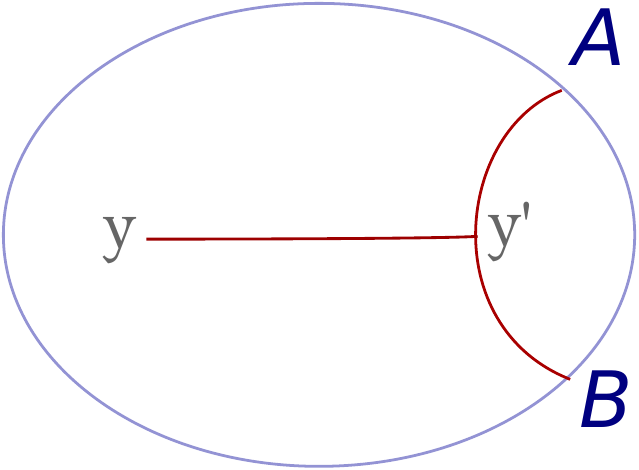}
    \end{flushright}
  \end{minipage}\hfill
  \begin{minipage}[c]{0.35\textwidth} 
    \caption{   Bulk geodesic prescription for computing the $\langle \Phi^{(0)} A B \rangle$ correlator.} 
  \end{minipage}
\end{figure}


\noindent More precisely, 
\bea
\Phi_{\g_{\!{}_{AB}}} (y) &= &\frac{2 \mathcal{B}^{-1}_{AB}}{|x_{12}|^{\D_A +\D_B}} \int_{\g_{\!{}_{AB}}} d\l \, e^{-\l \D_{AB}} \, G_{bb} (y, y'(\l))\non \\
&= & \frac{2}{ \mathcal{B}_{AB}}  \int_{\g_{\!{}_{AB}}} d\l \, G_{bb}(y, y'(\l)) G_{b\p} ( y'(\l),x_1) G_{b\p} ( y'(\l),x_2)\label{gwdr3pf}
\eea
The reason for the second equality is that the bulk-to-boundary propagator $G_{b\p}$, when evaluated between a point on the $\g_{\!{}_{AB}}$ geodesic and one its endpoints, reads 
 
 \be
 G_{b\p} ( y'(\l),x_1) = \frac{e^{- \l \D_A}}{|x_{12}|^{\D_A}} \;, \;\;\;\;\;\;\;\;G_{b\p} ( y'(\l),x_2) = \frac{e^{ \l \D_B}}{|x_{12}|^{\D_B}}  \label{Gbp}
 \ee
 Note that the three-point function $\Phi_{\g_{\!{}_{AB}}}$ is is different, at the relevant order in the $1/N$ expansion, from the full three-point correlator of the bulk field and two boundary operators, in that the bulk field correlator $\langle \Phi A B \rangle$ also receives double-trace contributions, which are essential in restoring bulk microcausality \cite{Kabat:2011rz}.

One can similarly use the relation between the bulk field and the boundary OPE to compute the four-point conformal partial wave

\be
\mathcal{W}_\O (x_i) =  \langle A(x_1) B(x_2) \mathbb{P}_\O \, C(x_3) D(x_4) \rangle \label{cbdef}
\ee
where $\mathbb{P}_\O$ denotes the projector onto the conformal family of the operator $\O$. This is given by

\be
\mathcal{W}_\O (x_i) = \frac{4\,\mathcal{B}_{AB}^{-1}\mathcal{B}_{CD}^{-1}}{|x_{12}|^{\D_A +\D_B}|x_{34}|^{\D_C +\D_D}}  \int_{\g_{\!{}_{AB}}} d\l\, e^{-\l \D_{AB}} \int_{\g_{{}_{CD}}} d\l' \, e^{-\l'\D_{CD}}\langle \Phi^{(0)}(y(\l)) \Phi^{(0)}(y'(\l')) \rangle \label{cbgwd}
\ee
Using the fact that $\langle \Phi^{(0)} \Phi^{(0)} \rangle$ is  the bulk-to-bulk propagator in AdS and the identity \eqref{Gbp}, the  geodesic Witten diagram prescription of \cite{Hijano:2015zsa} follows

\bigskip

\begin{figure}[h]
  \begin{minipage}{0.45\textwidth}
  \begin{flushright}
    \includegraphics[height=2.8cm]{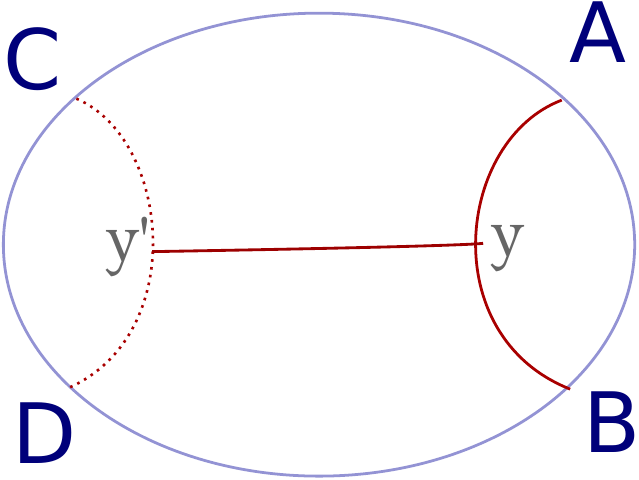}
    \end{flushright}
  \end{minipage}\hfill
  \begin{minipage}[c]{0.37\textwidth} 
    \caption{ Geodesic Witten diagram prescription for computing the conformal block, which follows from applying $(1.3)$ to the OPEs of the external operators. } 
  \end{minipage}
\end{figure}

\medskip

\noindent Note that one can also only use the relation between the bulk field and OPE once, to write the conformal partial wave as
\be
\mathcal{W}_\O (x_i) = \frac{2\,\mathcal{B}_{CD}^{-1}}{|x_{34}|^{\D_C +\D_D}}\int_{\g_{{}_{CD}}} d\l' \, e^{-\l'\D_{CD}} \, \Phi_{\g_{\!{}_{AB}}}(y(\l'))\label{cbvsbf}
\ee
i.e., the conformal partial wave is the Mellin  transform of the bulk field sourced on the geodesic from $A$ to $B$ and evaluated on the geodesic from $C$ to $D$. That  the conformal block could be written in this way was emphasized in \cite{Hijano:2015zsa}.  

The relations above between bulk fields and boundary correlators have been derived for operators that are all spacelike separated from each other (otherwise, \eqref{fggrels} does not hold). However, we may often  be interested in correlators where some of the boundary operators are timelike separated. In this case, we can still use  the geodesic Witten diagram prescription (which is defined in euclidean AdS) to compute correlation functions of the linearized bulk operator, provided we perform an  appropriate analytic continuation to Lorentzian signature at the end of the calculation.

The best setup to exemplify this is perhaps AdS$_2$, where the  only possible separation of the boundary operators is timelike. We start by studying the correlator $\Phi_{\g_{\!{}_{AB}}} = \langle A(t_1) B(t_2) \Phi^{(0)}(z,t)\rangle $. This correlator can be evaluated in a number of ways, the simplest of which is to use \eqref{2dbfope},
 which shows it is  equal to the four-point conformal partial wave with insertions at
$t_{1}, t_2$ and  

\be
t_3 =  t + z \;, \;\;\;\;\;\;\; t_4 =  t -z
\ee
We thus have 
\be
\langle A (t_1) B (t_2) \Phi^{(0)}(z,t)\rangle  = 2^{-\D} (t_{34})^{2\D_C} \mathcal{W}_\O (t_i)
\ee
where $\D_C=\D_D$, but $\D_{A,B}$ can be arbitrary. The one-dimensional conformal block\footnote{We are using the definition \eqref{defcb} for the conformal block, which is slightly different from that of \cite{Dolan:2011dv}.} can be found e.g. in \cite{Dolan:2011dv}
\be
\mathcal{G}_\O (t_i)= x^\D (1-x)^{\D_{CD}} {}_2  F_1 (\D-\D_{AB},\D+\D_{CD},2\D; x)\;, \;\;\;\;\;\;\; x= \frac{t_{12} t_{34}}{t_{13} t_{24}}
\ee
Setting for simplicity $t_1 = 0$ and $t_2=\infty$ on the plane,  the anharmonic ratio becomes

\be
 x = \frac{2 z}{t+z}
\ee
yielding
\be
\Phi_{\g_{\!{}_{AB}}} = \lim_{t_2 \r \infty} t_2^{2\D_B} \langle A(0) B(t_2) \Phi^{(0)}(z,t)\rangle   = t_4^{-\D_{AB}} \left(\frac{x}{2} \right)^\D {}_2F_1(\D-\D_{AB},\D,2\D,x)
\ee
It will be useful to write the end result in the following equivalent form\footnote{Obtained by using the hypergeometric  identity
\be
{}_2 F_1 (a,b,2b ; x) = (1-x)^{-a/2} {}_2F_1\left(\frac{a}{2},b-\frac{a}{2},b+\half ; \frac{x^2}{4x-4} \right)
\ee } 

\be
\Phi_{\g_{\!{}_{AB}}}  = \left( \frac{\D t^2}{t^2-\D t^2} \right)^{\D/2} \left( t^2-\D t^2\right)^{-\D_{AB}/2} {}_2 F_1 \left(\frac{\D-\D_{AB}}{2},\frac{\D+\D_{AB}}{2},\D + \half ; \frac{\D t^2}{\D t^2 -t^2} \right) \label{3pcads2}
\ee
Our proposal is that this correlator equals the bulk field souced on a geodesic in \emph{euclidean} AdS$_2$, which is then analytically continued to timelike separation. The euclidean solution for the bulk field  is given by \eqref{gwdr3pf}, specialised to  AdS$_2$. When the $A,B$ insertions are at $t_1 = 0$ and $t_2=\infty$ on the plane - which corresponds to sources at $\tau_E \r \pm \infty$ on the global AdS cylinder - the expression one obtains is \cite{Hijano:2015zsa}

\be
\Phi_{\g_{\!{}_{AB}}} (\rho,\tau_E) \sim  e^{- \D_{AB} \, \tau_E} \, (\cos\rho)^\D \; {}_2F_1\left(\frac{\D+\D_{AB}}{2},\frac{\D-\D_{AB}}{2},\D - \frac{d-2}{2} ;\cos^2\rho \right)  \label{phiglsc}
\ee
where $\rho, \tau_E$ are euclidean AdS$_2$ coordinates and  $d=1$. Transforming to  euclidean  Poincar\'e coordinates\footnote{ The map between global and Poincar\'e \emph{euclidean} AdS$_2$ is
$$e^{2\tau_E} = z^2 + t_E^2 \;, \;\;\;\;\;\;\; \cos^2\rho = \frac{z^2}{z^2 + t_E^2} $$
By contrast, the map between \emph{Lorentzian} Poincar\'e and global AdS$_2$ is \cite{Spradlin:1999bn}:  $t \pm z = \tan \half (\tau \pm \rho \pm \pi/2)$. }, the solution reads
\be
\Phi_{\g_{\!{}_{AB}}} (z,t_E)\sim  \left( \frac{z^2}{z^2 + t_E^2} \right)^{\frac{\Delta}{2}}\!\! (z^2+t_E^2)^{-\frac{\D_{AB}}{2}} {}_2F_1\left(\frac{\D+\D_{AB}}{2},\frac{\D-\D_{AB}}{2},\D - \frac{d-2}{2}  ; \frac{z^2}{z^2 + t_E^2} \right) \label{phipoinc}
\ee
Upon the analytic continuation $t_E \r i t$, this is identical to the correlator \eqref{3pcads2} for $d=1$.

 Note that these manipulations only hold when mapping \emph{euclidean} Poincar\'e AdS to \emph{euclidean} global AdS; in Lorentzian signature, timelike geodesics do not reach the boundary, and the coordinate transformation from global to Poincar\'e coordinates is quite different  (see   footnote) and  does not, in particular,  map $t = 0, \infty$   to $\tau \r \pm \infty $, or \eqref{phiglsc} to \eqref{phipoinc}.  

The relation \eqref{phipoinc} and its higher-dimensional analogues (obtained by simply replacing $t_E^2 \r |x|^2$) yield  the three-point function of one bulk and two boundary operators when the latter are inserted at $x_1=0$ and $ x_2 \r \infty$. It is also interesting to understand the dependence of $\Phi_{\g_{\!{}_{AB}}}$ on the positions $x_{1,2}$. The simplest way to proceed is by using \eqref{cbvsbf} and entails going through the steps of the original computation of \cite{fgg} of the four-point conformal blocks, which we review in  appendix \ref{cbdet}. Since the conformal block only  depends on the conformally invariant cross ratios of the four points $A,B,C,D$

\be
\rho = \frac{x_{14}^2 x_{23}^2}{x_{12}^2 x_{34}^2}\;, \;\;\;\;\;\;\;\;\; \eta = \frac{x_{13}^2 x_{24}^2}{x_{12}^2 x_{34}^2}
\label{crat}
\ee
we should be able to also express the  normalizable bulk field sourced on the $\g_{{}_{AB}}$ geodesic and evaluated along the $\g_{{}_{CD}}$ one in terms of  $\rho, \eta$. To find the expression, we rewrite
the bulk field solution \eqref{phipoinc} sourced at the center of euclidean AdS (which corresponds to taking $x_1 =0$ and $x_2 = \infty$) and evaluated along $\g_{{}_{CD}}$, with
\be
 \xi(\l) \equiv  \frac{z^2(\l)}{z^2(\l) + |x(\l)|^2} =   \frac{|x_{34}|^2}{|x_3|^2 \, (1+e^{-2\l}) + |x_4|^2 (1+ e^{2\l})}
\ee
and rewrite it in terms of the conformally invariant ratios \eqref{crat}. We obtain

 \be
 \Phi_n  (\rho, \eta, \s)= [\s(1-\s)]^{-\frac{\D_{AB}}{2}} \rho^{\frac{\D_{AB}}{2}} \xi^{\frac{\D+\D_{AB}}{2}}\, {}_2F_1 \left(\frac{\D+\D_{AB}}{2}, \frac{\D-\D_{AB}}{2}, \D; \xi  \right)
 \ee
with
\be
\xi(\s) = \left[ \frac{\rho}{\s} + \frac{\eta}{1-\s}\right]^{-1}\;, \;\;\;\;\;\;\; \s = \frac{1}{1+ e^{2\l}} \label{xis}
\ee 
This is related  to the conformal block, defined as

\be
\mathcal{W}_\O(x_i) = \frac{1}{|x_{12}|^{\D_A + \D_B} |x_{34}|^{\D_C+\D_D}} \left|\frac{x_{24}}{x_{14}}\right|^{\D_{AB}}  \left| \frac{x_{24}}{x_{23}}\right|^{\D_{CD}} \, \mathcal{G}_\O(\rho,\eta) \label{defcb}
\ee 
via

 \be
 \mathcal{G}_\O(\rho,\eta) = 2^{\D} \mathcal{B}_{CD}^{-1} \int_0^1 \frac{d\s}{\s(1-\s)}  \left(\frac{\s}{1-\s} \right)^{\frac{\D_{CD}}{2}} \Phi_n (\rho, \eta, \s) \label{relbfgb}
 \ee 
On the other hand, the relation \eqref{xis} together with the fact that $\xi = \cos^2\rho$ - where $\rho$ is the radial coordinate in global AdS -  identifies the radial depth (which is a function of the geodesic parameter) with a path in the plane of the CFT conformal ratios. Setting $x_1=0, x_2=\infty, x_4=1$ and $x_3$ arbitrary, this   can also be identified with the plane where the CFT lives. 

So far, we have been reviewing the relation between conformal partial waves  - which are Wightman functions - and normalizable bulk fields. Another four-point block of interest is the  euclidean amplitude, which can be computed using the shadow operator formalism of \cite{Ferrara:1972uq}

\be
\mathcal{W}_E (x_i) = \int d^d x \, \langle A(x_1) B(x_2) \O(x) \rangle \langle \O^\star (x) C(x_3) D(x_4) \rangle
\ee
This computation has also been performed in \cite{fgg}, with the result 

\bea
\mathcal{W}_E & \propto  & |x_{12}|^{-2\D_B} |x_{13}|^{-\D_{AB}-\D_{CD}} |x_{14}|^{-\D_{AB}+\D_{CD}} |x_{34}|^{\D_{AB}+\D_{CD}} \times   \label{ebsf}\\
& \times & \int_0^1 d \s\,  \s^{\frac{\D_{AB}+\D_{CD}}{2}-1} (1-\s)^{\frac{\D_{AB}-\D_{CD}}{2}-1}\, {}_2 F_1 \left[ \frac{\D^\star -\D_{AB}}{2}, \frac{\D-\D_{AB}}{2}, \frac{d}{2}, 1 - \left( \frac{\rho}{\s} + \frac{\eta}{1-\s}\right) \right] \non
\eea
Using hypergeometric identities, this can be turned into a particular linear combination  of the usual conformal block and its ``shadow'', which corresponds to the exchange of a fictitious operator of dimension $\D^\star = d-\D$ that has the same eigenvalue of the conformal Casimir as $\O$. Consistently with this, the euclidean amplitude above is related by a Mellin transform to an euclidean bulk field that is smooth in the center of AdS, but exhibits both normalizable and non-normalizable behaviour at infinity. 
Such a bulk field arises  if one naively tries to derive the OPE relation \eqref{fggope} from the shadow operator formalism, which naturally reproduces the \emph{time-ordered} three-point function in the CFT \cite{Ferrara:1972ay}.
The expression one obtains is very similar to \eqref{fggope}, except that the Bessel function 
 $J_\nu$ is replaced by the modified Bessel function $K_\nu \propto (I_\nu - I_{-\nu} )$, which contains contributions from the conformal families of both of operator of dimension $\D = \frac{d}{2} + \nu$ and a shadow operator of dimension $\D^\star =\frac{d}{2} - \nu $.  This particular linear combination is presumably required in order to have a finite contribution to the euclidean three-point function as the euclidean momentum $p_E \r \infty$.

\section{Bulk fields in BTZ \label{bfbtz}}

So far, we have discussed the relation between  the clasical bulk field in euclidean (global) AdS and boundary conformal blocks. We found that the bulk field that is normalizable at infinity (and thus has a source in the center of AdS - on a geodesic, in our case) is related via an integral transform to the usual conformal block, whereas the bulk field that is smooth in the interior (and thus has both normalizable and non-normalizable pieces at infinity) is related to the euclidean four-point conformal block, which is a particular linear superposition of the usual conformal block and its shadow. 

This relation has given us a means to identify the radial direction in the bulk - as probed by a \emph{local} bulk field - with the CFT plane, in particular with the plane of anharmonic ratios. Given that  every point inside (euclidean) vacuum AdS lies on some boundary-anchored geodesic, we can relate the bulk field at any point in the interior to a boundary conformal block as in \eqref{relbfgb}, by appropriately tuning the conformal ratios, all while  staying in the euclidean regime. Note that while the relation \eqref{relbfgb} between a free field in AdS and the boundary conformal block is purely kinematical - and  would thus hold in any CFT, including the Ising model - the requirement   that $\Phi^{(0)}$ represent a  \emph{local} bulk field in a weakly-coupled AdS quantum field theory necessitates that we work with a CFT with a large $N$ expansion, where correlation functions of light operators factorize. As was  nicely explained in \cite{ElShowk:2011ag}, the factorization property is naturally implemented by combining the light operators into a one higher-dimensional free field, and conformal symmetry of the CFT vacuum implies this field propagates in pure AdS.


What we are interested in, however,  is the bulk field in non-trivial backgrounds.  It is  expected that  the large $N$ expansion will  lead to  factorization of the CFT  correlators of light operators also around backgrounds whose energy scales with $N$, such as thermal backgrounds. In these cases, it is again natural to consider CFT operators that behave like approximately free and local bulk fields in a one-higher dimensional bulk. However, unlike around the CFT vacuum, where the construction of the bulk operators is basically fixed by conformal symmetry (and the wave equation they satisfy can be identified with the boundary Casimir equation, as we see from \eqref{fggrels}), around non-trivial backgrounds it is rather difficult to justify an expression for the emergent bulk field and the particular wave equation it satisfies by only using CFT input. For example, while one can certainly  use an HKLL-type construction for the bulk field, the procedure necessitates information from the bulk; in particular, it takes the radial direction as an input, which is not convenient for our purposes.  We may be able to understand the free bulk field in a nontrivial background as a  resummation of many vacuum diagrams, but so far this has not been done. 

Recently, there has been progress in understanding thermalization in two-dimensional CFTs with a large central charge, $c$, and factorization of the associated light correlators. More precisely, \cite{Fitzpatrick:2014vua,Fitzpatrick:2015zha} showed that pure states obtained by acting with a ``heavy'' operator on the CFT vacuum (i.e., an operator whose dimension $\D_H \propto c$) effectively look thermal when probed by ``light'' operators (whose dimension is $\O(1)$), and that correlation functions of the light operators in the heavy state factorize. These facts can then be used  provide the needed CFT justification for constructing the free bulk field in the dual gravitational background, which is nothing but the   three-dimensional BTZ black hole \cite{Banados:1992wn}. We will do so by using the methods of the previous section and the results of \cite{Fitzpatrick:2015zha} to relate  bulk fields in BTZ  to heavy-light Virasoro conformal blocks in the dual CFT, and  extract the meaning of the radial direction in the bulk from the CFT perspective. 

\subsection{Bulk fields and heavy-light conformal blocks}

Our starting point is the work of \cite{Fitzpatrick:2014vua,Fitzpatrick:2015zha}, who showed that in two-dimensional  CFTs with a large central charge, the correlation functions of light operators in pure states   $\O_H | 0 \rangle$ created by the insertion of a heavy operator look effectively thermal and factorize. 
One particularly neat result proven in  \cite{Fitzpatrick:2015zha} is that the heavy-light conformal block,  defined as the Virasoro conformal block where  the dimensions of the various operators scale as\footnote{In this section, we will be working in terms of holomorphic and anti-holomorphic dimensions $h,\bar h$; the same scaling relations below hold for the $\bar h$. We also use a slightly different normalization for the conformal blocks, in that  $\mathcal{G}_\O (w) \bar{\mathcal{G}}_\O (\bar w) = \langle A(0) B(\infty) \mathbb{P}_\O C(w,\bar w)D(1) \rangle_{global}$, which differs by a factor of $(1-w)^{-h_C - h_D}$ from that in the previous section. The Virasoro conformal block is similarly defined as
 $\mathcal{V}_\O (z) \bar{\mathcal{V}}_\O (\bar z) = \langle A(0) B(\infty) \mathbb{P}_\O C(z,\bar z), D(1) \rangle_{Vir}$, where the subscript indicates that we are summing over all the \emph{Virasoro} descendants of $\O$. }
\be
h_{A}, h_B \sim \O(c) \;, \;\;\;\;\;\;  h_C, h_D,  h_{AB}, h \sim \O(1)
\ee
reduces to the global conformal block in a set of conformally-transformed  coordinates $w,\bar w$

\be
\mathcal{V}_\O (h_i,h,z) = (2\pi i T_L)^{h_D-h}  (w'(z))^{h_C} \, \mathcal{G}_\O \left(\frac{h_{AB}}{2\pi i T_L}, h_C, h_D, h, w\right)\non 
\ee
\be
\bar{\mathcal{V}}_\O (\bar h_i, \bar h, \bar z) = (-2\pi i T_R)^{\bar h_D-\bar h}  (\bar w'(\bar z))^{\bar h_C} \, \bar{\mathcal{G}}_\O \left(-\frac{\bar h_{AB}}{2\pi i T_R}, \bar h_C,\bar  h_D,\bar h, w\right)
\label{vbgb}
\ee
which are given by\footnote{The particular signs in the exponents were obtained by relating the Poincar\'e coordinates \eqref{wec}  that correspond to an euclidean BTZ black hole to those corresponding to the CFT vacuum (which has $r_+ = -i, r_-=0$) keeping  $\phi, t_E$ fixed.  } 
\be
w(z) = z^{2\pi i T_L}\;, \;\;\;\;\;\;\;\bar{w}(\bar z) = \bar{z}^{- 2\pi i T_R} \label{wtoz}
\ee
with $T_{L,R}$ defined in \eqref{temp}. Note that the global blocks are effectively evaluated at imaginary values of the relative  conformal dimension, so these expressions must be defined via an appropriate analytic continuation. 

To show \eqref{vbgb}, \cite{Fitzpatrick:2015zha} used the well-known fact that when all operators involved are light, the Virasoro block simply reduces to the global conformal block. While this is no longer true when some of the external operators are heavy, \cite{Fitzpatrick:2015zha} found a trick - very special to two-dimensional CFTs - 
 that allowed them to absorb away the large matrix elements of the heavy operators with the Virasoro modes via the  conformal transformation \eqref{wtoz}. As a result, the computation of the heavy-light Virasoro block on the original CFT $z$ plane is  reduced to the computation of an all-light Virasoro block on the $w$ plane, which equals a global conformal block.  An interesting consequence of \eqref{vbgb}  is that now the heavy-light Virasoro block effectively satisfies a Casimir-type equation, due to its equality to a global conformal block in $w$ coordinates.

The equality \eqref{vbgb} can then be used to argue for thermality, as follows:  the expectation value of two light operators in the heavy background (which equals the heavy-light four-point function) reduces in the lightcone OPE limit (in which correlators are dominated by the exchange of the identity Virasoro block) to the global identity block in the $w$ background 

\be
\langle \O_H(\infty) \O_H (0) \mathbb{P}_{Id.} \, \O_L (z) \O_L (1) \rangle_{Vir} \sim \langle \O_{H'} (\infty) \O_{H'} (0) \mathbb{P}_{Id.}   \O_L (w) \O_L(1)  \rangle_{global}
\ee
The latter identity block is trivial, as it factorizes into the heavy and light contributions. Using the coordinate transformation \eqref{wtoz}, the light contribution  becomes the   thermal two-point function of the light operators in the CFT $(t,\phi)$ plane (with $z = e^{i (t-\phi)}$). One can similarly handle arbitrary correlation functions of light operators in the background of two heavy ones. In the lightcone OPE limit, they become
\be
\langle \O_H(\infty) \O_H (0)  \mathbb{P}_{Id.} \, \O_L (z_1) \O_L (z_2) \ldots \O_L (z_{2n})  \rangle_{Vir} \sim \langle \O_H(\infty) \O_H (0)  \mathbb{P}_{Id.} \, \O_L (w_1) \O_L (w_2)  \ldots \O_L (w_{2n})\rangle_{global} 
\ee 
Using large $N$, the RHS  factorizes into products $\langle \O_L \O_L \rangle_{w}$, each of which equals a thermal two-point function in the CFT plane. From this factorization, there immediately follows an HKLL-type prescription for the free bulk field $\Phi^{(0)}$ in BTZ, which is nothing but the vacuum one, but in the $w$ background 

\be
\Phi^{(0)} (z, w^+, w^-) = \int dw'^+ d  w'^- K_P(z,w^+,w^- | w'^+, w'^-)\, \O(w'^+, w'^-) \label{kabw}
\ee
where $K_P $ is the AdS$_3$ bulk-to-boundary propagator in Poincar\'e coordinates \cite{Hamilton:2006fh}. The  dependence on the particular BTZ background under consideration now comes from the fact that the map \eqref{wtoz} between the CFT coordinates $z$ and the $w$ is state-dependent.
The expression \eqref{kabw} can be brought to a more familar form by taking into account the identification \eqref{widentif} of the $w^{\pm}$ coordinates
to rewrite the boundary integral as being over over a single fundamental region covered by the $t,\phi$ coordinates, at the price of having to sum over the images $K_P^{(n)}$ of the propagator under $\phi \r \phi + 2 \pi n$. This produces the well-known propagator in the BTZ background, and is thus equivalent to the usual HKLL prescription, if we  carefully keep track  of the range of integration\footnote{When the bulk point is in the interior of the BTZ black hole, part of the support for the $w'$ integral will lie in the second asymptotic region of BTZ.}. 

Using the equivalence between the HKLL formula in vacuum AdS and the inverse Mellin transform of the OPE, we should be able to relate the bulk field \emph{in the $w$ plane } to the global conformal family of $\O$ in the $w$ plane. The latter can in principle  be re-expressed in terms the $z$-plane global conformal families of various multitrace operators constructed from $\O$ and the stress tensor, using the map between the $z$-plane and  $w$-plane Virasoro generators given in \cite{Fitzpatrick:2015zha}. It would be interesting to  relate this to the diagrammatic expansion of the heavy-light Virasoro block recently developed in \cite{Fitzpatrick:2015foa}. 


In the following, we will rather be interested in the \emph{classical} bulk field in the black hole background, which is given by 
 $\langle \Phi^{(0)} \O_H \O_H \rangle$. This can be extracted using the relation between the heavy-light Virasoro block on the CFT plane and the global conformal block on the $w$ plane, which is the geodesic integral/Mellin transform of the classical bulk field

\be
\langle \O_H (\infty) \O_H (0) \mathbb{P}_{\O} \O_L(z) \O_L(1) \rangle_{Vir} \sim \langle \O_H (\infty) \O_H (0) \mathbb{P}_{\O} \O_L(w) \O_L (1) \rangle_{global} \sim  \int_{\g_{{}_{LL}}}  d\l \, e^{-\l \D_{LL}}\, \Phi_{\g_{{}_{LL}}}
\ee 
Everything we have said in the previous section concerning the relation between bulk fields and boundary conformal blocks will still hold, with the CFT $z$-plane being replaced by the $w$-plane. 
The bulk field satisfies the vacuum Casimir equation in the $w$ plane, which becomes the wave equation in BTZ. It corresponds to the field sourced on a geodesic in $w$ coordinates. Upon transforming to black hole coordinates, this is the bulk field sourced on a backreacted geodesic. As before, the above relation will allow us to identify the radial BTZ coordinate with the conformal ratios  \emph{on the $w$ plane}. 

It is interesting to ask whether the procedure of extracting $\langle \Phi^{(0)} \O_H \O_H \rangle$ from the conformal block is state-independent - after all,  it only appears to involve ``undoing'' the OPE integral of the light operators inside the block, without touching the heavy operators. 
 It would be interesting to have a diagrammatic understanding of our procedure of ``undoing'' the OPE that would confirm state-independence.

\subsection{Properties of the bulk solution in BTZ}

In this subsection, we study solutions to the free wave equation in BTZ and find their corresponding counterparts in Poincar\'e and in global coordinates. We will be concentrating on the BTZ black hole with mass
$M=r_+^2+r_-^2$ and angular momentum $J=2r_+r_-$, whose metric is given by\footnote{We changed conventions as $t \r -t$ with respect to those in \cite{Maldacena:1998bw}. } 

\begin{equation}
ds^2=-\frac{(r^2-r_+^2)(r^2-r_-^2)}{r^2}dt^2+\frac{r^2dr^2}{(r^2-r_+^2)(r^2-r_-^2)}
+r^2\left(d\phi-\frac{r_+r_-}{r^2}dt\right)^2
\label{btzmetric}
\end{equation}
where $\phi\sim \phi+2\pi$. The BTZ metric is locally, but not globally, AdS$_3$. This means that it can be put in the form
\be
ds^2 = \frac{dz^2 + d w^+ dw^-}{z^2}
\ee
where the map between the Schwarzschild and Poincar\'e coordinates is \cite{Maldacena:1998bw}

\be
w^\pm = \left( \frac{r^2-r_+^2}{r^2-r_-^2}\right)^{\half} \, e^{2\pi T_\pm (\phi \mp t)} \;, \;\;\;\;\;\;\; z = \left( \frac{r_+^2-r_-^2}{r^2-r_-^2}\right)^{\half} \, e^{\pi T_+ (\phi-t) + \pi T_- (\phi+t)} \label{wtphilor}
\ee
The left/right temperatures are defined as 

\be
T_L = T_+ = \frac{1}{2\pi} (r_+ + r_-) \;, \;\;\;\;\;\;\; T_R = T_- = \frac{1}{2\pi} (r_+ - r_-)\label{tpm}
\ee
Note that even before taking into account the $\phi$ identification, the $\phi,t$ coordinates only cover the right diamond of the $w^\pm$ plane - thus, the corresponding state is thermal, with temperatures $T_{L,R}$. The $\phi$ identification acts on the $w^\pm$ coordinates as 

\be
w^\pm \sim e^{4\pi^2 T_\pm} w^\pm \label{widentif}
\ee
and a fundamental region is shown in the picture below. 

\bigskip

\begin{figure}[h]
  \begin{minipage}{0.45\textwidth}
  \begin{flushright}
    \includegraphics[height=4.2cm]{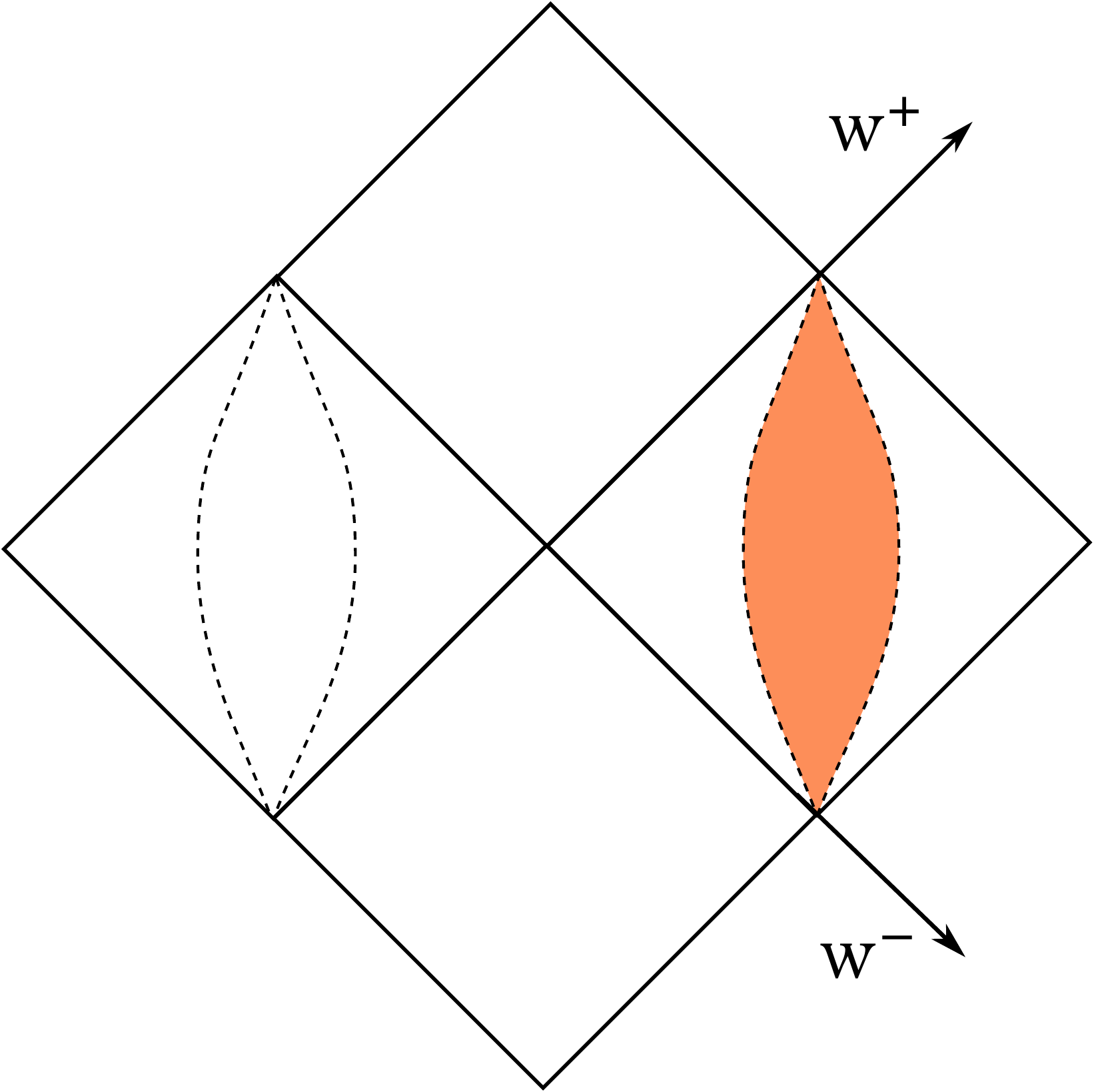}
    \end{flushright}
  \end{minipage}\hfill
  \begin{minipage}[c]{0.36\textwidth} 
    \caption{The CFT $(t,\phi)$ coordinates only cover  the right Rindler wedge of the $w$ plane. The shaded region represents   a particular fundamental region in the $w$ plane covered by the $\phi, t$ cooordinates.  } 
  \end{minipage}
\end{figure}

\medskip

\noindent
 For our applications, it will be useful to also consider the Euclidean BTZ black hole, whose metric is given by \eqref{btzmetric} with the replacement $t = - i t_E $.  The metric is still real if  $r_-$ is purely imaginary, and the solutions to the wave equation stay exactly the same up to some extra factors of $i$. Note that the left/right temperatures \eqref{tpm} are now complex conjugates of each other.
  The euclidean black hole is related via the following 
 coordinate transformation to Euclidean Poincar\'e  with coordinates $z, w,  \bar w$ 

\be
w = \left( \frac{r^2-r_+^2}{r^2-r_-^2}\right)^{\half} \, e^{(r_++r_-)(\phi + i t_E)} \;, \;\;\;\;\;\;\; z = \left( \frac{r_+^2-r_-^2}{r^2-r_-^2}\right)^{\half} \, e^{r_+ \phi + i r_- t_E} \label{wec}
\ee
The quotient now acts as $w \sim e^{4\pi^2 T_+} w$ and the fundamental region is an annulus. 

Next, we consider the wave equation  for a scalar field of mass $m$ in this background. The equation of motion $(\Box-m^2)\Phi=0$  is separable. Writing

\be \Phi(t,r,\phi)=e^{-i\omega
  t+i \k\phi}R(\rho)\;, \;\;\;\;\;\;\; \rho=r^2
  \ee
the radial equation is 
\begin{equation}
\frac{d}{d\rho}\left[(\rho-r_+^2)(\rho-r_-^2)\frac{d}{d\rho}R(\rho)\right]
-\left(\frac{m^2}{4}-\frac{(\omega^2-\k^2)(\rho-r_-^2)+(\k r_+-\omega
    r_-)^2}{4(\rho-r^2_+)(\rho-r^2_-)}\right)R(\rho)=0,
\end{equation}
which is a Fuchsian equation with three regular singular points at
$\rho=r_\pm^2$ and $\rho=\infty$. The critical exponents\footnote{Defined via  $R(\rho) \sim (\rho -r_i^2)^{\half\theta_i}(1+\ldots)$ near each singular point.} at each point
are 

\be
\frac{1}{2}\theta_+ = \pm i \frac{r_+\omega -r_-\k}{2(r_+^2-r_-^2)}\;, \;\;\;\;\;\;
\frac{1}{2}\theta_- =\pm  i \frac{r_-\omega-r_+\k}{2(r_+^2-r_-^2)},\qquad
\frac{1}{2}\theta_\infty = - \frac{1}{2}(1\pm\sqrt{1+m^2})
\ee
Introducing the quantity  
\be
\xi = \frac{r_+^2-r_-^2}{r^2-r_-^2}
\ee
the normalizable solution to the wave equation reads
\be
\Phi_n = e^{- i \om t + i \k \phi} (1-\xi)^{\frac{i(\om r_+ - \k r_-)}{2(r_+^2-r_-^2)}} \xi^{\D/2} \, {}_2 F_1 \left(\frac{\D}{2} + \frac{i(\om-\k)}{2(r_+-r_-)},\frac{\D}{2} + \frac{i(\om+\k)}{2(r_++r_-)} , \D, \xi \right) \label{nsolbtz}
\ee
The non-normalizable solution $\Phi_{nn}$ is simply obtained from the above via the replacement $\D \r 2-\D$. At the horizon, the solution needs to be analytically continued to a function of $1-\xi$. The singular points of the solution occur at $\xi =0$ ($r \r \infty$), $\xi =1$ ($r = r_+$) and $\xi \r\infty$ ($r = r_-$). The Euclidean solution is obtained via the analytic continuation $t \r - i t_E$.

The above bulk field solution in the euclidean black hole background is  just the coordinate transformation 
of the bulk field sourced on a geodesic at the center of euclidean global AdS
\be
\Phi_n^g= e^{- (h^g_{AB} +\bar h^g_{AB}) \tau_E + i (h^g_{AB}-\bar h^g_{AB}) \varphi}  \, {}_2 F_1 \left(\frac{\D}{2} + h^g_{AB}, \frac{\D}{2} -\bar h^g_{AB},\D,\cos^2\rho \right) (\cos  \rho)^\D (\sin\rho)^{ h^g_{AB}- \bar h^g_{AB}} \label{phigl}
\ee
with arbitrary energy and angular momentum parametrized by $h^g_{AB}, \bar h^g_{AB}$, as above. The change of coordinates between euclidean Schwarzschild/Poincar\'e/global AdS$_3$ coordinates reads

\be
\frac{r_+^2-r_-^2}{r^2-r_-^2} = \frac{z^2}{z^2 + w\bar w} =\cos^2\rho\;, \;\;\;\;\;\; e^{r_-\phi+ i r_+ t_E} = \sqrt{\frac{w}{\bar w}}=e^{-i\varphi} \;, \;\;\;\;\;\;\; e^{r_+ \phi+i r_- t_E} = \sqrt{z^2+w \bar w}=e^{\tau_E} \label{cootr}
\ee
Note the linear relation between the  black hole and the global AdS time and angular coordinates. Such a relation only exists in \emph{euclidean} signature; in Lorentzian signature the relation between global and Poincar\'e coordinates is quite different. Under this change of coordinates, \eqref{nsolbtz} is equivalent to \eqref{phigl}, provided we identify
\be
h^g_{AB} \r - \frac{i(\om+\k)}{4\pi  T_L} \;, \;\;\;\;\;\;\;\bar h^g_{AB} \r  \frac{i(\om-\k)}{4\pi  T_R} \label{hhbrep}
\ee
Noting that $\half (\om \pm \k)$ are the left/right conformal dimensions in the original CFT $(t,\phi)$ plane, we find that the relation \eqref{vbgb} between the global conformal block and the heavy-light Virasoro block is consistent with the relation between the global conformal block and  the bulk field \eqref{phigl} in pure AdS and the fact that BTZ is obtained from AdS$_3$  via a simple  change of coordinates.

It is also interesting to ask what does the solution in global AdS that is smooth in the interior become after the map \eqref{cootr} to the black hole background. The smooth solution  is (for $h^g_{AB} > \bar h^g_{AB}$)

\bea
\Phi_s & = & \Phi_n - \frac{ \Gamma (\Delta ) \Gamma \left(h^g_{AB}-\frac{\Delta }{2}+1\right) \Gamma \left(-\bar h^g_{AB}-\frac{\Delta }{2}+1\right)}{\Gamma (2-\Delta ) \Gamma \left(h^g_{AB}+\frac{\Delta }{2}\right) \Gamma \left(\frac{\Delta }{2}-\bar h^g_{AB}\right)}\, \Phi_{nn} =  e^{- (h^g_{AB} +\bar h^g_{AB}) \tau_E + i (h^g_{AB}-\bar h^g_{AB}) \varphi}  \times  \\
& \times & \, \xi^{\frac{\D}{2}} (1-\xi)^{\frac{h^g_{AB}-\bar{h}^g_{AB}}{2}}  \frac{\G(1-\bar h^g_{AB} - \frac{\D}{2})\G(1+ h^g_{AB} - \frac{\D}{2})}{\G(1+h^g_{AB}-\bar h^g_{AB}) \G(1-\D)} \, {}_2 F_1 \left(\frac{\D}{2} +h^g_{AB} , \frac{\D}{2} - \bar h^g_{AB}, 1+ h^g_{AB} - \bar h^g_{AB}, 1-\xi \right)\non
\eea
Introducing  the tortoise coordinate 

\be
r_\star = -\frac{1}{2(r_+^2-r_-^2)}  \left[r_+ \ln \frac{r+r_+}{r-r_+} - r_- \ln \frac{r+r_-}{r-r_-} \right] \; \underset{r \r r_+}{\approx} \; \frac{r_+}{2(r_+^2-r_-^2)}  \ln (1-\xi)
\ee
and using the replacement \eqref{hhbrep}, then we find that $\Phi_s \propto e^{-i \om (t+r_\star)}$, i.e. it is the purely ingoing solution. 
 A simple hypergeometric transformaton

\bea
{}_2 F_1 \left(\frac{\D}{2} +h^g_{AB} , \frac{\D}{2} - \bar h^g_{AB}, 1+ h^g_{AB} - \bar h^g_{AB}, 1-\xi \right) &= &  \\
&& \hspace{-6cm} = \xi^{ \bar h^g_{AB}- \frac{\D}{2}} \, {}_2 F_1 \left(\frac{2-\D}{2} - \bar h^g_{AB} , \frac{\D}{2} - \bar h^g_{AB}, 1+ h^g_{AB} - \bar h^g_{AB}, 1-\xi^{-1} \right) \non
\eea
brings it to the same  form as the integrand appearing in $\mathcal{W}_E$, provided we set  $d=2$, $  h^g_{AB}=  \bar h^g_{AB}$.  Thus, $\mathcal{W}_E$ is the Mellin transform of the analytic continuation of the bulk field with ingoing boundary conditions at the horizon\footnote{Given that the effective global conformal  dimensions \eqref{hhbrep} are imaginary,  this statement should be accompanied by an appropriate analytic continuation, obtained e.g. by  giving $h_{AB}^g$ a small real part.}, upon the replacement \eqref{hhbrep}. That such a relationship exists is perhaps not surprising, in view of the fact that i) the finiteness of the euclidean correlator selects a unique combination of the usual and shadow conformal block, which can be written  as the Mellin transform \eqref{ebsf} of an euclidean bulk field that is smooth in the center of AdS, and ii) the BTZ bulk field with ingoing boundary conditions at the horizon is always related by the coordinate transformation \eqref{wtphilor} and analytic continuation to the smooth euclidean bulk field.

%
%

\section{CFT analysis of the bulk solution \label{cfti}}

We have proposed that by looking inside a conformal block we can extract the bulk field at some radial position in the interior. So far, we have mapped the bulk field in BTZ to a bulk field in AdS, with the radial positions identified as \eqref{cootr}. In turn, the integral of the bulk field along a geodesic  in global AdS is related to the global conformal block in the $w$ plane as in \eqref{relbfgb}, with the identification between the global AdS radial coordinate (evaluated along the geodesic, with parameter $\l/ \s$) and the boundary anharmonic ratios given in \eqref{xis}. Using \eqref{cootr}, this gives us the following identification between the BTZ radial coordinate and the conformal ratios \emph{in the $w$ plane}

\be
\xi^{-1} (\s) = \frac{r^2(\s)-r_-^2}{r_+^2-r_-^2} =\frac{\eta}{1-\s} + \frac{\rho}{\s} \label{radconf}
\ee
This can be translated to the $\l$ affine parameter via  $\s = (1+e^{2\l})^{-1}$. The conformal ratios are

\be
\rho = \frac{|w_{14}|^2 |w_{23}|^2}{|w_{12}|^2 |w_{34}|^2}\;, \;\;\;\;\;\; \eta = \frac{|w_{13}|^2 |w_{24}|^2}{|w_{12}|^2 |w_{34}|^2} 
\ee
In turn, the global conformal block on the $w$ plane is related via \eqref{vbgb} to the heavy-light Virasoro block in the original CFT coordinates, and the conformal ratios in the $w$ plane are related to specific functions of the $z$-plane conformal ratios and $T_{L,R}$; however,  we will not need to make use of this relation in the present section - we will just be working on the $w$ plane all along.  

Thus, the relation \eqref{radconf} lets us identify the radial coordinate of the bulk field with the CFT conformal ratios (both depend on the free parameter $\s$) and \eqref{relbfgb} allows us to relate the bulk field to the boundary conformal block. In the following, we discuss each of these separately.

\subsection{Mapping the singular points}

We would now like to use \eqref{radconf} 
to identify the singular points of the radial wave equation (located at $r_{\pm}, \infty$) with special points in the plane of conformal ratios of the CFT. We will fix the operator insertions at $w_1=0, w_2 = \infty, w_4=1$, which are fixed points under the map \eqref{wtoz} and will vary $w_3 \equiv w, \; \bar w_3 \equiv \bar w$ in the complex plane. The euclidean regime will correspond to $\bar w = w^\star$, whereas in the Lorentzian regime the two are real and  unrelated. 

The mapping of the singular points is as follows:
\bi
\item $r \r \infty $  corresponds to $ \rho \; \mbox{or} \; \eta \r \infty $, which can be naturally achieved by $w_3 \r w_4$. In this case, the $\g_{CD}$ geodesic on which we are evaluating the bulk solution lies very close to the AdS boundary, and thus it is natural that it probes the respective singular point.

\item $ r \r r_+ $ can be achieved  for either $\rho \r 0, \eta \r 1, \s \r 0 $, as long as $\rho$ approaches zero faster than $\s$, or for $\eta \r 0, \rho \r 1, \s \r 1 $. This  singular point can thus be reached from within the euclidean regime by sending $w_3 \r w_2 $ or $w_3 \r w_1 $.
\item $ r \r r_- $ requires that both $ \rho, \eta \r 0$.  This corresponds to the double lightcone OPE limit $w_{31}^2 \r 0 $ and $ w_{32}^2 \r 0$, which is an intrinsically Lorentzian limit. 
\ei 
In terms of the coordinates $w,\bar w$, the mapping of the singular points can be summarized as follows

\begin{center}
\begin{tabular}{c|c}
$r$ & $w, \bar w$  \\ \hline
$\infty$ & $w, \bar w \r 1$ \\
$r_+$ & $w,\bar w \r 0, \; \l \r -\infty$
   \\
 $r_-$ & $ w \r 0, \bar w \r \infty$ 
\end{tabular}
\end{center}
where in the last two rows we have  an additional option where we interchange $0$ and $\infty$, as can be seen from the relation between $r$ and $w$

\be
\xi(\l) = \frac{(1-w)(1-\bar w)}{1 + e^{2\l} + w \bar w(1 + e^{-2\l}) } = \frac{r_+^2 - r_-^2}{r^2(\l) - r_+^2} \label{xil}
\ee
We can access different bulk regions by varying $\s$ (or, equivalently, $\l$). The region outside the horizon has $\xi^{-1} >1$, and can be reached for  $\rho,\eta$ belonging to the euclidean regime  $\bar w = w^\star$. The region inside the horizon has $\xi^{-1}  <1$, which is an intrinsically Lorenzian regime\footnote{One can easily see that the minimum of the expression \eqref{radconf} is $(\sqrt{\rho} + \sqrt{\eta})^2$ and that in the euclidean regime, this quantity is always greater than one.}. In terms of the Poincar\'e coordinates $z,w^\pm$, this region corresponds to having $w^2(\l) = w^+ (\l) w^-(\l) < 0$ for some $\l$.  Along the $\g_{{}_{CD}}$ geodesic, we have

\be
\frac{w^2(\l) }{z^2(\l)} = \frac{(w_3 e^{-\l} + w_4 e^\l)^2 }{w_{34}^2} = \frac{(w_3^+ e^{-\l} + w_4^+ e^\l)(w^-_3 e^{-\l} + w^-_4 e^\l) }{(w^+_3-w^+_4)(w^-_3-w^-_4)}
\ee
We assume that the boundary endpoints $w_{3,4}$ are spacelike separated; also, since we would like the geodesic to reach the boundary, we require that $w^2(\l) >0$ as $\l \r \pm \infty$, which implies that $w_3^+ w_3^- >0 $ and $w_4^+ w_4^- >0$. Without loss of generality, we take $w_3^\pm >0$; then $w_4^\pm <0 $ if we would like there to exist a range of $\l$ for which $w^2(\l) <0$. This region will have finite range if  $w_3^+/w^+_4 \neq  w^-_3/w^-_4$.  Thus, we find that in order for the conformal block to explore the region behind the horizon, we need the points $w_{3,4}$ to belong to different quadrants of the $w^\pm$ plane. Using the Lorentzian map \eqref{wtphilor} between the $w$ plane and the  CFT coordinates $(t,\phi)$, we note that one of the insertions in the conformal block will lie  outside the region of the $w$ plane that is covered by the original CFT coordinates; in the spacetime picture, this translates into the fact that one  endpoint of the $\g_{{}_{CD}}$ geodesic will belong to the  second  asymptotic region of the eternal BTZ black hole. Our construction is reminiscent of the work of \cite{Kraus:2002iv} on probing the interior of the eternal BTZ black hole with geodesics anchored on its two opposite boundaries.

Note however that the BTZ black hole we are considering corresponds to a \emph{pure} state in a single-sided CFT, which is rather different from the thermofield double state dual to the eternal black hole \cite{Maldacena:2001kr} and  is \emph{not} expected to have a second asymptotic region. Despite this, we find that in order to explore the region behind the horizon using the conformal block, we need to continue it analytically  in the emergent $w$ plane,  and this effectively produces a second asymptotic region. Thus, in our construction, the region behind the horizon cannot be explored by CFT observers with access to a single boundary, and consequently they cannot  predict the experience of an infalling observer that falls behind the horizon.  Note also that the appearance, via analytic continuation, of the left quadrant of the $w$ plane is effectively  a geometrization of  the entanglement between the light degrees of freedom around the heavy state and the rest of the CFT, as the expected doubling  of the light operator algebra occurring \cite{Papadodimas:2013jku}  is represented as geometric entanglement between the left and right Rindler wedges of the $w$ plane.


\subsection{Mapping the monodromies} 
 
So far, we have identified the radial plane in BTZ with the plane of conformal ratios in the conformally transformed background $w,\bar w$. As we showed in section \ref{crrgwd}, the  conformally-invariant classical bulk field solution \eqref{phigl} is related via a Mellin transform to the global conformal block in the $w$ plane

 \be
\mathcal{ G}_\O(\rho,\eta) = 2^{\D} \mathcal{B}_{CD}^{-1} \int_{-\infty}^\infty d\l  \, e^{-\l \D_{CD}} \Phi_n (\rho, \eta, \l) \label{genglrel}
 \ee 
Here, $\Phi_n$ is given by \eqref{phigl} and the global block is evaluated for \emph{general} dimensions $h^g_{A,B}, \bar h^g_{AB}$ of two of the external operators ($C,D$ are still taken to be scalars), which slightly generalizes\footnote{This can be shown as follows: The relation \eqref{fggrels} between the $CD$ OPE and the free bulk field $\Phi^{(0)}$ still holds, as $C,D$ are scalars. Inserting this into the conformal block, all we need to show is that $\langle \Phi^{(0)} A B \rangle$ equals \eqref{phigl} for arbitrary $A,B$. The bulk field is given the HKLL expression e.g. in \cite{Hamilton:2006fh}\be
\Phi^{(0)}(z,x,t) = \frac{\D-1}{\pi} \int_{y'^2+t'^2 < z^2} dy' dt' \left( \frac{z^2 - y'^2- t'^2}{z} \right)^{\Delta-2} \O(x+i y', t + t') = \G(\D) \, z^\D \sum_m \frac{ (-1)^m z^{2m}}{\G(m+\D) m!} \, \p_+^m \p_-^m \O(x^+,x^-)   \non
\ee 
where $x^\pm = x \pm t$.
 } the all-scalar relation we discussed in section  \ref{crrgwd}.  

%


 
Both the LHS and the integrand on the RHS of \eqref{genglrel} are given by hypergeometric functions\footnote{The holomorphic part of the  global block  in two dimensions is: $(1-w)^h {}_2 F_1 (h-h_{AB},h+h_{CD}, 2h, 1-w)$.}, whose singular points have been identified according to the table in the previous subsection, and whose monodromies around the various singular points are known. We could  thus try to match the monodromies on the two sides of \eqref{genglrel}. 

Both hypergeometric functions have definite monodromies around $r \r \infty/ w \r 1$.   As $w \r 1$, the monodromy of the LHS is simply $2h$. The monodromy of the RHS comes from the singular point $\xi \r 0$ and is $\l$-independent (thus, we can factor it out of the integral) and also equals $2h$, as can be seen from \eqref{nsolbtz} and \eqref{xil}. Note that it is important, for the monodromies to match, that we work on the CFT $w$ -  \emph{plane}; working on the $z$ plane  would multiply the monodromy of the conformal block by additional factors of $T_{L,R}$, which would not match the radial bulk monodromy. In other words, the radial plane in  BTZ is really identified  with the $w$ plane, and not the original CFT $z$ plane. It is probably  factorization that singles out  the $w$ coordinates.

For the remaining two singular points, neither side of  \eqref{genglrel}  has definite monodromy, but rather is a known superposition of functions with definite monodromy. The monodromies of the conformal blocks on the LHS generally depend on $h_{CD}$; however, the only way that we could have the RHS monodromy depend on this relative dimension is to have the geodesic parameter $\l$ itself transform as the singular points are taken around each other. This is reminiscent of the comments of \cite{Fitzpatrick:2015dlt} on the interplay between  the monodromies of the conformal block  and the geodesic Witten diagram representation of the latter. It would be very interesting to understand the connection between bulk and boundary monodromies more precisely; we leave a careful investigation of this issue to future work.

The monodromy around  the outer horizon  appears to be particularly difficult to match, as  reaching this singular point  involves both a limit on $w$ and on $\l$, but the geodesic  parameter is integrated over. It may also be simpler to look instead at the relation between the bulk field with  ingoing boundary conditions at the horizon and the euclidean conformal block, as the former at least  has definite monodromy around the horizon. 
Note also that the global conformal block  has branch cuts in the $w$ plane. As discussed in e.g. \cite{Fitzpatrick:2015zha}, these branch cuts are unphysical and should disappear from the full correlators.

\section{Discussion}

In this article, we analysed the relation between bulk fields in AdS and conformal blocks in the dual CFT by expanding on the previous works of \cite{Ferrara:1971vh,fgg,Hijano:2015zsa}.  Using \cite{Ferrara:1971vh}, we pointed out a new interpretation of the bulk field as a CFT operator that encodes the full conformal family of a given primary, i.e.  the particular linear combination of primaries and descendants  that appears in the OPE of two other operators. We also proposed the that the bulk field can be extracted from the boundary OPE via an inverse Mellin transform with respect to the relative conformal dimension. While this relation is purely kinematic, when  applied to CFTs with a large $N$ expansion it can extract the  physical, approximately local dual bulk field at any point inside AdS. Also, it  identifies the radial direction in AdS with the plane of CFT  conformal ratios and the Casimir  equation on the boundary with the bulk wave equation.


What is more interesting is that the relation between bulk fields and boundary conformal blocks continues to hold for BTZ black hole backgrounds dual to heavy pure states in the dual CFT \cite{Hijano:2015qja}. This is a consequence of  two facts: i) that heavy-light conformal blocks in CFT$_2$  equal the global conformal block in a set of conformally transformed coordinates, $w$, and ii) that BTZ is a quotient of pure AdS, and thus the corresponding bulk fields are related by a simple coordinate transformation. 
This relation allows us to write the bulk fields in BTZ as the inverse Mellin transform of the boundary heavy-light Virasoro block, and thus to identify the radial BTZ plane with the  plane of CFT conformal ratios.   

Whereas in the work of \cite{Fitzpatrick:2015zha}, the conformally transformed coordinates - $w$ - were used mostly as a trick for simplifying the calculation of the Virasoro block, from the point of our work - which is  concerned with the radial behaviour of the bulk field in BTZ - the $w$ plane appears to be physically important, for several reasons: i) factorization of the thermal CFT correlators, which was necessary in order to have a free bulk field, occurs in the $w$ coordinates; ii) the heavy-light Virasoro conformal block, via its equality to the $w$-plane global block, satisfies a Casimir equation \emph{on the $w$ plane}, which becomes the wave equation for the free bulk field. This wave equation is background-dependent, as expected, because the map between the original  CFT coordinates and the $w$ ones itself is state-dependent; iii) the radial direction in the bulk is identified with the conformal ratios \emph{in the $w$ plane}; in particular, exploring the region behind the horizon of the black hole necessitates analytically continuing in the $w$ plane to a region that is not covered by the original CFT coordinates; iv) finally, the monodromies of the bulk field around the singular points of the bulk wave equation are captured by the $w$ - \emph{plane} monodromies. In a certain sense, the role of the $w$ plane is to  geometrize the entanglement between the light degrees of freedom around the heavy state and the rest of the CFT, by representing it as entanglement of the $w$ vacuum when seen from the point of view of the CFT Rindler-type coordinates. However, as we just saw, the $w$ - plane appears to take on a life of its own when we ask questions that are natural from the bulk point of view.

Note that, while in the case of pure AdS, the relation between the bulk field and the boundary OPE  was entirely kinematic, this is not true for the relation between the bulk field in BTZ and the associated boundary CFT operators, as the latter relies essentially on factorization in the $c \r \infty$ limit. It would be very interesting to study the boundary representation of the bulk operator in BTZ in  light of the diagrammatic representation  of the heavy-light Virasoro block \cite{Fitzpatrick:2015foa} and understand how to construct the bulk field in  this nontrivial background from summing vacuum diagrams. While the end result will of course agree with the bulk field correlators we discussed, the advantage of such a construction   would be that the $w$ plane - which encodes information about the horizons - should emerge from this procedure. This  may allow us to e.g. probe the interior of the BTZ black hole without having to invoke geodesics that stretch to the second, unphysical boundary of non-eternal BTZ.  

It would also be very interesting to extend the connection between bulk fields and boundary conformal blocks to higher dimensions. The reason that studying singular points of the bulk wave equation and its associated monodromies may be a fruitful avenue is that the behaviour of free bulk fields in the vicinity of a horizon appears to be universal across many different black holes, as revealed by the  works on hidden conformal symmetries \cite{Castro:2010fd,Chen:2010bh}.

\subsection*{Acknowledgements}

We are grateful to Jan de Boer, Pawel Caputa,  Sheer El-Showk and  Nick Halmagyi for enlightening discussions, and especially to  Alejandra Castro for 
collaboration in the early stages of this project. The research of M.G. has been supported in part by the Swedish Research Council  grant number 113410213. B.C.C. acknowledges support from the FACEPE grant APQ-0051-1.05/15 and PROPESQ-UFPE.

\appendix

\section{Details of the conformal block calculations \label{cbdet}}

%

Our starting point will be  \eqref{cbgwd}, which is also the expression that \cite{fgg}   encounters after summing over all intermediate operators $\O(p)$ in the conformal family 
 (see  eqn. (28) in their paper). In the notation of \cite{fgg}, the conformal partial wave reads
\be
\hat{\mathcal{W}} (x_i) = \int_0^1 du f_{AB} (u) \int_0^1 dv  f_{CD}(v)\, \l_+^{- \D}\, {}_2 F_1 \left(\frac{\D+1}{2},\frac{\D}{2}, \D - \frac{d-2}{2}, \l_+^{-2}\right) \label{gwdfgg}
\ee
where we  defined the ``hatted'' conformal partial wave to be the partial wave \eqref{cbdef} with some overall factors  stripped
\be
\mathcal{W}_\O (x_i) = \frac{\mathcal{B}_{AB}^{-1}\mathcal{B}_{CD}^{-1}}{|x_{12}|^{\D_A +\D_B}|x_{34}|^{\D_C +\D_D}} \;  \hat{\mathcal{W}}_\O (x_i)
\ee
The $u,v$ integrals simply represent the geodesic/Schwinger parameter integrals; the relation between the parameters $u,v$ and our previous parameters $\l, \l'$ in \eqref{cbgwd} are

\be
u = \frac{e^{2\l}}{1+e^{2\l}} \;, \;\;\;\;\;\;\; v = \frac{e^{2\l'}}{1+e^{2\l'}}
\ee
The function $f_{AB}$ reads
\be
f_{AB} (u) = \left(\frac{u}{1-u}\right)^{\frac{\D_{AB}}{2}-1}
\ee
and $f_{CD}(v)$ takes an analogous form. The hypergeometric function in \eqref{gwdfgg} is nothing but the bulk-to-bulk propagator in AdS$_{d+1}$  and  $\l_+ (u,v,x^i)$ is the invariant distance in AdS between points on the two geodesics 
\be
\l_+ = \frac{z^2(u) + z'^2(v) + [x(u)-x'(v)]^2}{2 z(u) z'(v)}
\ee
where
\be
z(u) = \sqrt{u(1-u) x_{12}^2} \;, \;\;\;\;\;\; x(u) = x_2 + u \, x_{12} 
\ee
and similarly for $z'(v), x'(v)$. The end expression for the invariant distance is

\be
\l_+ =  \frac{x_{24}^2 (1-u)(1-v) + u(1-v) x_{14}^2 + v (1-u) x_{23}^2 + u v \, x_{13}^2}{2 \sqrt{ uv(1-u)(1-v) x_{12}^2 x_{34}^2 }}
\ee
The expression \eqref{gwdfgg} holds at spacelike separation, and the signature is Lorentzian. 

Following \cite{Hijano:2015zsa}, we would like to re-express \eqref{gwdfgg} as an integral over the bulk field sourced on the $\g_{\!{}_{AB}}$ geodesic. Note that the numerator in $\l_+$ is linear in $u$.  To obtain the bulk field souced on $\g_{AB}$, we  integrate over $u$. This is possible by  using the Barnes integral representation of the hypergeometric

\be
{}_2 F_1 (a,b,c,z) = \frac{1}{2\pi i} \frac{\G(c)}{\G(a) \G(b)} \int_{-i\infty}^{i\infty} d\tau \frac{\G(a+\tau) \G(b+\tau) \G(-\tau) }{\G(c+\tau)} (-z)^\tau \label{barnesh}
\ee
The $u$ integral reduces to

\be
 \int_0^1 du \, f_{AB} (u) \, \l_+^{-\D -2 \tau}(u) = \mathrm{y}^{-\D -2 \tau} \int_0^1 du \, u^{\frac{\D + \D_{AB}}{2} + \tau -1} (1-u)^{\frac{\D - \D_{AB}}{2} + \tau -1} \left( 1-  \mathrm{x} \, u\right)^{-\D -2\tau}
\ee
where

\be
\mathrm{x} = \frac{v\, (x_{23}^2- x_{13}^2) + (1-v) (x_{24}^2-x_{14}^2)}{v\, x_{23}^2 + (1-v)\, x_{24}^2}\;, \;\;\;\;\;\;\; \mathrm{y} = \frac{v\, x_{23}^2 + (1-v)\, x_{24}^2}{2 \sqrt{v(1-v) x_{12}^2 x_{34}^2}}
\ee
The $u$ integral is precisely the integral representation of the  hypergeometric  ${}_2 F_1 (2\tau + \D, \frac{\D+\D_{AB}}{2} + \tau, 2\tau + \D, \mathrm{x})$, but this function simply reduces to $ (1-\mathrm{x})^{-( \frac{\D+\D_{AB}}{2} + \tau)} $. The expression for the conformal block now becomes

\be
\hat{\mathcal{W}}_\O (x_i) = \int_0^1 dv \, f_{CD}(v) \,  \mathrm{y}^{-\D} (1-\mathrm{x})^{- \frac{\D+\D_{AB}}{2}} \, \frac{1}{2\pi i} \frac{\G(\D - \frac{d-2}{2})}{\G(\D/2) \G((\D+1)/2)}\, I_\tau
\ee
where

\be
I_\tau =  \int_{-i \infty}^{i\infty} d\tau \frac{\G(\frac{\D+1}{2}+\tau) \G(\frac{\D}{2}+\tau) \mathcal{B} (\frac{\D+\D_{AB}}{2}+\tau, \frac{\D-\D_{AB}}{2}+\tau) \G(-\tau) }{\G(\D - \frac{d-2}{2} + \tau) } \, [-4 \mathrm{y}^2 (1-\mathrm{x})]^{-\tau}
\ee
The integrand can be simplified by using the identity

\be
\G(z) \G(z + \half) = 2^{1-2z} \sqrt{\pi}\, \G(2z)
\ee
and the expression for the conformal block becomes

\be
\hat{\mathcal{W}}_\O (x_i) = \mathcal{B}_{AB}\int_0^1 dv \, f_{CD}(v) \,  \mathrm{y}^{-\D} (1-\mathrm{x})^{- \frac{\D+\D_{AB}}{2}}\, {}_2 F_1 \left(\frac{\D+\D_{AB}}{2},\frac{\D-\D_{AB}}{2},\D-\frac{d-2}{2}, (2 \, \mathrm{y})^{-2}(1-\mathrm{x})^{-1}\right)
\ee
where we have again used \eqref{barnesh}. Plugging in the expressions for $\mathrm{x}, \mathrm{y}$, this yields 
\bea
\hat{\mathcal{W}} (x_i)& = & 2^\D \, \mathcal{B}_{AB} |x_{12}|^\D |x_{34}|^\D\int_0^1 dv \tilde{f}_{CDO} (v) [x_{24}^2 (1-v) + v \, x_{23}^2]^{-\half(\D-\D_{AB})}  [x_{14}^2 (1-v) + v \, x_{13}^2]^{-\half(\D+\D_{AB})} \non \\
& \times & {}_2 F_1 \left(\frac{\D-\D_{AB}}{2},\frac{\D+\D_{AB}}{2},\D-\frac{d-2}{2},\frac{v (1-v) x_{12}^2 x_{34}^2}{(x_{24}^2 (1-v) + v \, x_{23}^2)(x_{14}^2 (1-v) + v \, x_{13}^2)} \right) \label{cbvint}
\eea
 The integrated object is the bulk field $\langle \Phi^{(0)} AB \rangle$ sourced on the $\g_{{}_{AB}}$ geodesic and evaluated on the $\g_{{}_{CD}}$ one. 
 The function $\tilde f_{CDO}(v)$ in \eqref{cbvint} is defined as

\be
\tilde f_{CDO}(v) = f_{CD} (v)\, [v(1-v)]^{\frac{\D}{2}}
\ee
Note that the integrand is not (yet) over the conformally invariant cross-ratios, although the integrated object is the bulk field. We can again use the Barnes representation of the hypergeometric to analyse the $v$ integral. Leaving aside the overall factors and the Gamma functions, the $v$ integral reads

\be
\int_0^1 dv \, v^{\frac{\D+\D_{CD}}{2} + \tau -1} (1-v)^{\frac{\D-\D_{CD}}{2} + \tau -1} (1-v x)^{-\frac{\D-\D_{AB}}{2} - \tau } (1-v y)^{-\frac{\D+\D_{AB}}{2} - \tau }
\ee
where 
\be
x= 1 - \frac{x_{23}^2}{x_{24}^2} \;, \;\;\;\;\;\;\; y = 1 - \frac{x_{13}^2}{x_{14}^2}
\ee
Naively, this looks like the integral representation of the Appell hypergeometric function, 

\be
F_1 \left(\frac{\D+\D_{CD}}{2} + \tau; \frac{\D-\D_{AB}}{2} + \tau, \frac{\D+\D_{AB}}{2} + \tau; \D+2 \tau; x, y\right)
\ee
However, note that we are in the special case in which $\g = \b + \b'$, where the Appell is known to reduce to a single-variable hypergeometric. This can be easily shown by performing a coordinate transformation of the form

\be
v = \frac{(1-a) \s}{1-a \s}
\ee
and then choosing $a= x/(x-1)$ (or $a= y/(y-1)$). In any case, we do not perform the $\s$ integral, obtaining

\bea
\hat{\mathcal{W}}_\O (x_i) &= & 2^\D \mathcal{B_{AB}} |x_{12}|^\D |x_{34}|^\D |x_{24}|^{\D_{AB} + \D_{CD}} |x_{14}|^{-\D-\D_{AB}} |x_{23}|^{-\D-\D_{CD}} \times \label{fullw}\\
& \times &\int_0^1  d\s f_{CDO} (\s) (1-\s (1-\zeta))^{- \frac{\D + \D_{AB}}{2}} \, {}_2 F_1 \left(\frac{\D+\D_{AB}}{2}, \frac{\D-\D_{AB}}{2}, \D - \frac{d-2}{2}; \left[ \frac{\rho}{\s} + \frac{\eta}{1-\s}\right]^{-1} \right) \non
\eea
where the conformal ratios have been defined in \eqref{crat}. This agrees with (30) in FGG and leads to the relation \eqref{relbfgb}.

\end{document}